# Suppression of amplitude-to-phase noise conversion in balanced optical-microwave phase detectors

Maurice Lessing,[1,2] Helen S. Margolis,[1] C. Tom A. Brown,[2] Patrick Gill,[1] and Giuseppe Marra[1*]

[1]*National Physical Laboratory, Hampton Road, Teddington, Middlesex, TW11 0LW, UK*
[2]*SUPA, School of Physics and Astronomy, University of St Andrews, St Andrews, Fife, KY16 9SS, UK*
[*]*giuseppe.marra@npl.co.uk*

**Abstract:** We demonstrate an amplitude-to-phase (AM-PM) conversion coefficient for a balanced optical-microwave phase detector (BOM-PD) of 0.001 rad, corresponding to AM-PM induced phase noise 60 dB below the single-sideband relative intensity noise of the laser. This enables us to generate 8 GHz microwave signals from a commercial Er-fibre comb with a single-sideband residual phase noise of $-131$ dBc Hz$^{-1}$ at 1 Hz offset frequency and $-148$ dBc Hz$^{-1}$ at 1 kHz offset frequency.

## 1. Introduction

Low-noise microwave signals are required for many applications such as radar systems, telecommunications, long baseline interferometry and precision spectroscopy. Lasers locked to high-finesse Fabry-Pérot cavities made of ultra-low expansion glass are amongst the most stable oscillators available in any region of the electromagnetic spectrum. When a frequency comb is locked to such an ultra-stable optical oscillator it acts as an optical frequency divider, transferring the stability of the optical reference to the microwave domain [1]. The phase noise of such photonic oscillators has been shown to be comparable to the best microwave oscillators available [2, 3].

One of the main challenges in this approach to the generation of low-noise microwave signals is the conversion of the stable optical pulse train into a stable electronic pulse train. Saturation effects in photodiodes limit the power of the extracted microwave signal and introduce excess phase noise due to amplitude-to-phase noise (AM-PM) conversion [4, 5], which converts intensity fluctuations of the laser into phase fluctuations of the microwave signal. The AM-PM conversion coefficient $\alpha$ for commercial photodiodes is typically between 0.1 rad and 2 rad. This is too high to achieve effective suppression of the typical relative intensity noise (RIN) of commercially available mode-locked lasers. For example, for the laser used in [6] the equivalent phase noise level would, at best, be $-117$ dBc Hz$^{-1}$ at 1 Hz offset frequency and $-150$ dBc Hz$^{-1}$ at 100 kHz offset frequency.

Several techniques have been developed to address these issues. The AM-PM conversion coefficient depends on the optical energy per pulse and is observed to vanish at particular pulse energies. After characterizing this dependence for commonly used photodetectors Zhang *et al.* reported that it should be possible to maintain a value of $\alpha$ below 0.03 rad over long periods of time by tuning the system to operate near one of these vanishing points [4]. This value of $\alpha$ corresponds to an AM-PM induced phase noise 30 dB below the RIN of the laser. At high offset frequencies the phase noise is also limited by the power of the microwave signal available from the saturated photodiode and can be improved using repetition-rate multiplication techniques based on cascaded Mach-Zehnder interferometers. By combining repetition-rate multiplication with operation at a pulse energy where $\alpha$ was close to zero, Haboucha *et al.* were able to generate 12 GHz signals with a residual phase noise of $-162$ dBc Hz$^{-1}$ at 100 kHz offset frequency [7]. The recent development of high-power, high-linearity modified unitravelling carrier photodiodes has further improved the phase noise floor at high offset frequencies. Using such photodiodes together with a repetition-rate-multiplied Ti:sapphire comb Fortier *et al.* generated 10 GHz microwave signals with a residual phase noise of approximately $-172$ dBc Hz$^{-1}$ at 100 kHz offset frequency [8]. However the disadvantage of the repetition-rate multiplication technique for generation of microwave signals with high long term stability is the temperature sensitivity of the Mach-Zehnder interferometers. Hence, although the long term stability of microwave signals extracted using photodiodes has been improved by optical power stabilization via feedback to the pump laser current [9], such power stabilization techniques have not yet been combined with repetition-rate multiplication.

One simple setup that can simultaneously provide both ultra-low noise microwave signals and subfemtosecond long term stability is the balanced optical-microwave phase detector (BOM-PD) [10, 11]. Using this device, Jung and Kim were able to generate 8 GHz signals from an optical pulse train with a residual phase noise of $-133$ dBc Hz$^{-1}$ at 1 Hz offset frequency and $-154$ dBc Hz$^{-1}$ at 5 kHz offset frequency [10]. The long-term stability was similar to the results reported by Zhang *et al.* [9] but without the need for optical power stabilization. The BOM-PD has the additional advantage that it can be operated over a range of repetition rates, whereas the repetition-rate multiplication technique requires exact matching between the repetition rate and the path length difference in the Mach-Zehnder interferometer. However, until now it has not been shown that the BOM-PD can suppress AM-PM conversion efficiently. Jung and Kim measured $\alpha$ for their BOM-PDs to be between 0.06 rad and 0.3 rad depending on the offset frequency [11], values which are very similar to those for photodiodes and which correspond to an AM-PM induced phase noise between 11 dB to 24 dB below the single-sideband RIN of the laser. In this work we report a modified BOM-PD design that results in a significantly improved AM-PM conversion coefficient of 0.001 rad.

## 2. Balanced optical-microwave phase detector and residual phase noise measurement

A schematic of the BOM-PD setup is shown in Fig. 1. The BOM-PD synchronizes an 8 GHz dielectric-resonator oscillator (DRO) to the 80$^{th}$ harmonic of the 100 MHz repetition rate of an Er-fibre comb. After the erbium-doped fibre amplifier the optical power is approximately 40 mW and spans a bandwidth of about 30 nm centred at 1545 nm. The DRO signal is applied to a unidirectional travelling-wave high-speed phase modulator within a Sagnac interferometer. The phase error $\theta_e$ between this signal and the pulse train in the optical domain is detected by converting it into an intensity imbalance between the two output ports of the Sagnac interferometer. The powers $P_1$ and $P_2$ at the two output ports are given by $P_1 = P_{avg} \cos^2(\Delta\Phi/2)$ and $P_2 = P_{avg} \sin^2(\Delta\Phi/2)$ where $P_{avg}$ is the average output power of the Sagnac interferometer and $\Delta\Phi$ is the phase difference between the counter-propagating beams. To lock the DRO frequency to the repetition rate of the optical pulse train requires the intensity at the two outputs to be balanced. This is achieved by introducing a non-reciprocal phase shift of $\pi/2$ between the two counter-propagating beams using two Faraday rotators and a quarter waveplate. $P_1$ and $P_2$ are detected using a balanced photodetector and the difference between the two photocurrents is fed into a transimpedance amplifier to yield an error signal proportional to $\theta_e$.

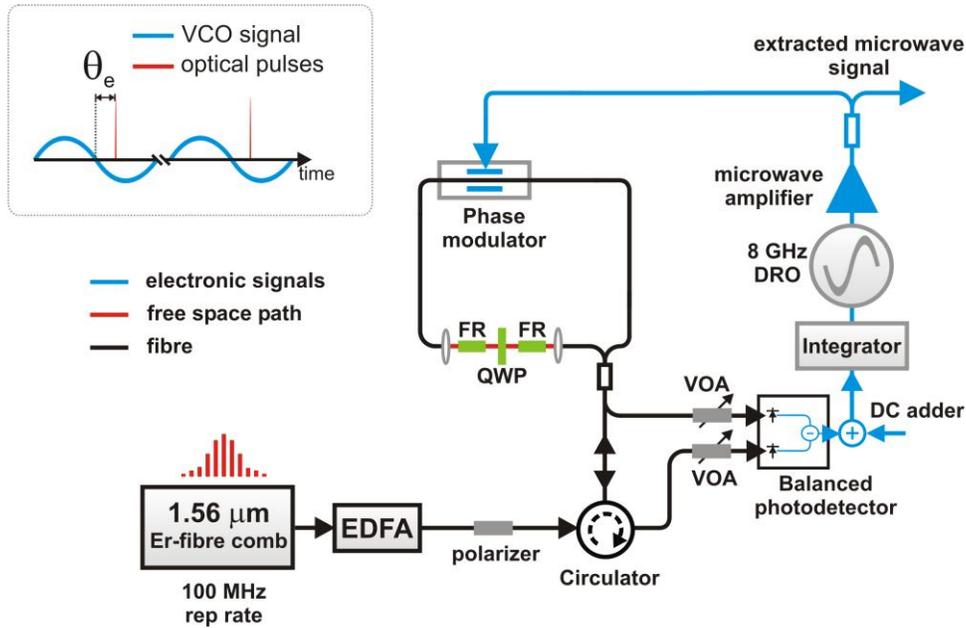

Fig. 1. Schematic of BOM-PD. EDFA: erbium-doped fibre amplifier; FR: Faraday rotator; QWP: quarter waveplate; DRO: dielectric-resonator oscillator; VOA: variable optical attenuator. All fibres after the polarizer are polarization maintaining. The inset shows the phase error $\theta_e$ between the microwave signal and the pulse train in the optical domain.

In order to achieve high AM-PM suppression we have made two key changes to the BOM-PD setup described by Jung and Kim [10]. Firstly, we have introduced variable optical attenuators (VOAs) at each output of the Sagnac interferometer. These compensate for the loss in the optical circulator and ensure that the balanced intensity condition is met, which is crucial for effective suppression of AM-PM conversion. Secondly, we add a DC voltage to the error signal to compensate for unwanted offsets in the loop filter electronics. Since only fairly coarse adjustments could be made using our VOAs, this DC voltage is also used to fine tune the balanced condition.

In order to measure the residual phase noise of the extracted microwave signal we built two nominally identical BOM-PDs. BOM-PD1 is locked to the 80$^{th}$ harmonic of the comb repetition rate while BOM-PD2 is used as an ultra-sensitive phase detector fed with the 8 GHz signal from the locked system as shown in Fig. 2.

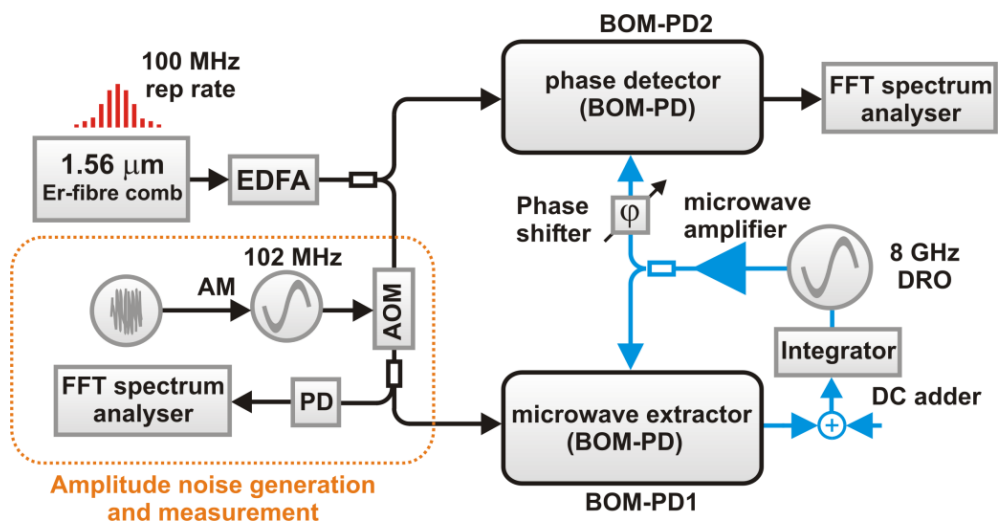

Fig. 2. Phase noise measurement set-up. EDFA: erbium-doped fibre amplifier; AOM: acousto-optic modulator; PD: photodiode; DRO: dielectric-resonator oscillator, BOM-PD: balanced optical-microwave phase detector. Also shown is the setup used to generate and measure amplitude noise for the experiments described in section 3.

In Fig. 3 we plot the single-sideband (SSB) phase noise versus the offset frequency in the range 1 Hz to 100 kHz. From the data shown, it can be seen that residual phase noise levels of −131 dBc Hz$^{−1}$ at 1 Hz offset frequency and −148 dBc Hz$^{−1}$ at 1 kHz offset frequency are achieved. Even though these experiments use a commercial Er-fibre frequency comb with a relatively high RIN (−116 dBc Hz$^{−1}$ at 10 Hz and −128 dBc Hz$^{−1}$ at 100 kHz), this performance is only slightly worse (by 2 dB at 1 Hz offset frequency and 6 dB at 5 kHz offset frequency) than the best results reported to date from a BOM-PD [10], which were achieved using a specially designed Er-fibre laser [12] with RIN approximately 17 dB lower [13]. Furthermore the optical power incident on each photodiode in our measurement is only 400 µW, almost an order of magnitude below that used by Jung and Kim [11]. The RF power we apply to each phase modulator is approximately 16 dBm, which yields a phase-to-voltage conversion coefficient of $K_d = 30$ rad V$^{−1}$ and a shot noise level of −154 dBc Hz$^{−1}$.

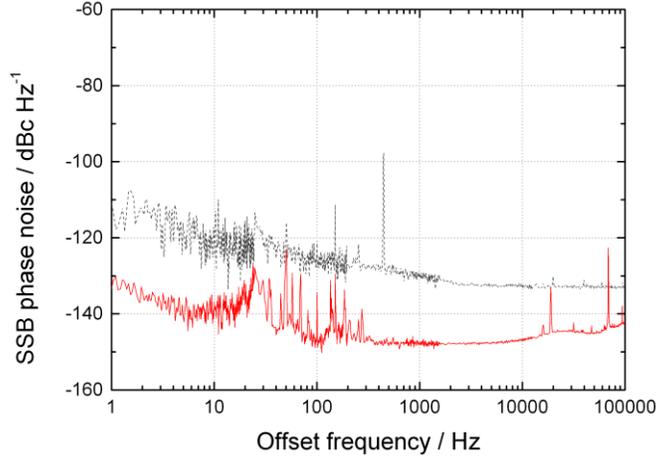

Fig. 3. Residual phase noise results. Solid red: BOM-PD. To estimate the contribution from a single system, 3 dB has been subtracted from the measured phase noise. Dashed grey: typical photodiode performance, shown for comparison.

The excess noise in the region between 10 Hz and 300 Hz is attributed to vibrations of the waveplate mount. By increasing the locking bandwidth, which is limited to several hundred kHz in our current setup, we expect to reduce the noise at high offset frequencies to the shot noise limit. Since the phase detector BOM-PD2 is unlocked for these measurements, the AM-PM suppression in this system cannot be fine-tuned using the DC adder, and there is still some propagation of AM noise through to the phase noise measurements. This issue could be resolved in future experiments either by locking both BOM-PDs and comparing them using a carrier suppression phase noise measurement system [14] or by using a fully optimized optomechanical design for the VOA mounts.

### 3. Measurement of the AM-PM conversion coefficient

The set-up for the AM-PM conversion coefficient measurement is shown in Fig. 2. We use an acousto-optic modulator to amplitude modulate the laser light incident on the locked BOM-PD with white noise up to 50 kHz. The RIN of the modulated laser signal is measured with a photodiode and then compared with its contribution to the phase noise of the generated 8 GHz microwave signal which is measured by the phase detector.

As shown in Fig. 4a, we achieve an AM-PM suppression of up to 60 dB. Of this, around 37 dB is achieved using the VOAs, whilst the remaining 23 dB is obtained using the DC adder. At lower offset frequencies and at offset frequencies above 50 kHz the suppression is limited by the residual noise of BOM-PD2 which is used to measure the phase noise of BOM-PD1. From these results, $\alpha$ can be calculated using the formula $L_{RIN}(f) = RIN_{SSB}(f) + 20\log(\alpha)$. As illustrated in Fig. 4b, we achieve a value of $\alpha = 0.001$ rad, an improvement by a factor of approximately 60–300 compared to the results of Jung and Kim [11]. As expected, $\alpha$ is constant over the range of offset frequencies where the measurement is not limited by the background noise of the phase noise measurement system. This is in contrast to the frequency-dependent $\alpha$ reported in [11].

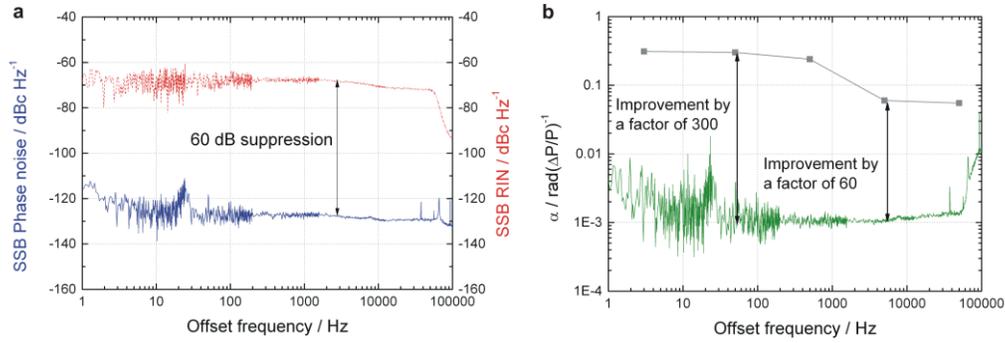

Fig. 4. a) AM-PM suppression results. Dashed red: RIN of modulated laser signal; solid blue: phase noise measured by phase detector BOM-PD2. b) AM-PM conversion coefficient. Solid green: calculated AM-PM conversion coefficient; grey squares: AM-PM conversion coefficient from Jung and Kim [11].

## 4. Conclusion

In summary, we demonstrate how to reduce the AM-PM conversion coefficient of a BOM-PD to $\alpha = 0.001$ rad, corresponding to AM-PM induced phase noise 60 dB below the single-sideband relative intensity noise of the laser. This enables us to generate 8 GHz microwave signals from a commercial Er-fibre comb with a residual phase noise of $-131$ dBc Hz$^{-1}$ at 1 Hz offset frequency and $-148$ dBc Hz$^{-1}$ at 1 kHz offset frequency. These phase noise levels are only 2–6 dB higher than the best reported results from a BOM-PD, which were achieved using a specially designed laser with very low RIN and an optical power approximately ten times higher. With the level of AM-PM suppression we have demonstrated, the RIN contribution to the phase noise of microwave signals extracted from optically referenced frequency combs with optimized RIN would become negligible. For example, for the laser used by Zhang *et al.* [4], its contribution would be below $-190$ dBc Hz$^{-1}$ between 1 Hz and 1 MHz, significantly below other major noise contributions.

**Acknowledgements**

This work is supported by the UK National Measurement System and the European Metrology Research Programme (EMRP). The EMRP is jointly funded by the EMRP participating countries within EURAMET and the European Union. ML acknowledges support from EPSRC through the Doctoral Training Centre in Optics and Photonics Technologies.